\documentclass[preprint,aps,amsmath,nofootinbib,tightenlines,floatfix]{revtex4}
\usepackage{graphicx, epsfig}

\newcommand{\eeq}{\end{equation}}
\newcommand{\beq}{\begin{equation}}
\newcommand{\bea}{\begin{eqnarray}}
\newcommand{\eea}{\end{eqnarray}}

\begin{document}

\setlength{\unitlength}{1mm}

\title{Dark Matter in the Left Right Twin Higgs Model}
 
\author{Ethan M. Dolle\footnote{edolle@physics.arizona.edu}, 
Shufang Su\footnote{shufang@physics.arizona.edu}}
\affiliation{Department of Physics, University of Arizona, Tucson, AZ 85721
}

\begin{abstract}
In the left-right twin Higgs model, one of the neutral Higgses is a natural candidate for WIMP dark matter.  We analyzed the dark matter relic density in this framework and identified regions of parameter space that provide the right amount of dark matter.  We also studied the dark matter in the more general inert Higgs doublet model in which the mass splittings between the dark matter and  other particles do not follow the relations in the left-right twin Higgs model.
\end{abstract}
 
 
\maketitle


\section{Introduction}
\label{sec:intro}
 
The Standard Model (SM) of particle physics has been extremely successful in explaining the data from particle physics experiments.  It is, however,  not satisfactory when it comes to cosmology.  One of the puzzles that remain is the amount of cold dark matter in the Universe, which has been determined precisely from WMAP\cite{wmap}: 
\begin{equation}
\Omega_{\rm DM}h^2=0.112 \pm 0.009.
\end{equation}
None of the SM particles can be a good candidate for the dark matter.  The existence of the dark matter itself points to new physics beyond the SM.  Weakly interacting massive particles (WIMPs) are promising dark matter candidates among various proposals in the literature.  Given the natural size of its interactions (weak interaction) and natural value of its mass around the TeV scale, the WIMP relic density is around the observed amount of dark matter without much fine tuning. 

Recently, the twin 
Higgs mechanism has been proposed as a solution to the little hierarchy
problem\cite{twinhiggsmirror, twinhiggsleftright, susytwinhiggs}.  
Higgses emerge as  pseudo-Goldstone bosons once a global symmetry is
spontaneously broken.  Gauge and Yukawa interactions that break  the 
global symmetry give masses to the Higgses, with the leading order being 
quadratically divergent.  Once an additional  discrete symmetry (twin symmetry) is imposed, the leading one-loop
quadratically divergent terms respect the global symmetry.  Thus they do not 
contribute to the Higgs masses.  The Higgs masses do obtain one-loop logarithmically
divergent contributions, resulting in masses around the electroweak scale when the 
cutoff is around 5$-$10 TeV.  

The twin Higgs mechanism can be implemented in left-right models 
with the discrete symmetry being identified with left-right 
symmetry\cite{twinhiggsleftright}.  
In the left-right twin Higgs (LRTH) model, several physical Higgs bosons remain
after the spontaneous symmetry breaking. In particular, there is an ${\rm SU}(2)_L$ Higgs doublet 
$\hat{h}$, which only couples to the gauge boson sector due to some additional discrete symmetry.  The lightest particle in its neutral components is stable, therefore a good candidate for  WIMP dark matter. 

The dark matter candidate in the left-right twin Higgs model is similar to the one in the inert Higgs doublet model (IHDM)\cite{inerthiggs}.  In the IHDM, one of the Higgs doublets plays the role of the SM Higgs doublet, which couples to both the fermion matter sector and the gauge boson sector.  The other Higgs doublet, however, only couples to the gauge boson sector similar to $\hat{h}$ in the LRTH model.   Therefore, it could contain a dark matter candidate. 

There are some differences between these two models though, which are relevant for the dark matter studies.  In the IHDM, the mass for the SM Higgs boson is a free parameter, which can be anywhere between 100 GeV up to a very large value.  The SM Higgs mass in the LRTH model, on the other hand, is more constrained.  It depends weakly on other model parameters and is typically between 160 $-$ 180 GeV.  
The Higgs potential in the LRTH model is also more constrained compared to that of the IHDM.   The dominant contributions to the Higgs potential in the LRTH model come from the one-loop Coleman-Weinberg (CW) potential.  The quartic couplings due to gauge interactions are about the size of $g^4/(16 \pi^2)$, much smaller than typical quartic couplings in the IHDM, in which the natural size is ${\cal O}(1)$.
In the LRTH model, however, there are extra heavy states and interactions, which lead to new features in the relic density analysis that are absent in the IHDM.

The relic density analysis of the dark matter in the IHDM has been studied in Ref.~\cite{inerthiggsrelic}.  Their results agree with our analysis.  We have furthermore identified regions of  parameter space that have previously been overlooked in Ref.~\cite{inerthiggsrelic}.  In addition, due to the extra particles and 
interactions that appear in the LRTH model, there are new regions that could produce the right amount of relic density, which are otherwise absent in the IHDM. 

This paper is organized as follows. In Sec.~\ref{sec:model}, we briefly introduce the LRTH model, emphasizing the Higgs sector.   In Sec.~\ref{sec:darkmatter}, we identify the dark matter candidate and introduce the parameters that are involved in the analysis.  In Sec.~\ref{sec:analysis}, we present in detail our numerical results for the dark matter relic density.  We discuss two  dark matter mass regions: (A) low mass region, and (B) high mass region, and identify the regions of parameter space in which we obtain the right amount of dark matter relic density.  In Sec.~\ref{sec:conclusion}, we conclude.

\section{The left-right twin Higgs model}
\label{sec:model}

The LRTH model was first proposed in Ref.~\cite{twinhiggsleftright} and the details of the model as well as the Feynman rules, particle spectrum, and collider phenomenology have been studied in Ref.~\cite{twinhiggssugoh}.  Here we briefly introduce the model and focus our attention on the Higgs sector.

In the LRTH model proposed in Ref.~\cite{twinhiggsleftright},
the global symmetry is ${\rm U}(4)\times {\rm U}(4)$, with the
diagonal subgroup of ${\rm SU}(2)_L\times {\rm SU}(2)_R\times
{\rm U}(1)_{B-L}$ gauged. 
The twin symmetry which
is required to control the quadratic divergences of the Higgs mass is identified
with the left-right symmetry which interchanges L and R. For the
gauge couplings $g_{2L}$ and $g_{2R}$ of
${\rm SU}(2)_L$ and ${\rm SU}(2)_R$, the
left-right symmetry implies that $g_{2L}=g_{2R}=g_2$.

Two Higgs fields, $H$ and $\hat{H}$, are introduced and each
transforms as $(\textbf{4},\textbf{1})$ and
$(\textbf{1},\textbf{4})$ respectively under the global symmetry.
They can be written as
\begin{equation}
H=\left(
\begin{tabular}{c}
$H_L$\\
$H_R$
\end{tabular}
\right),\ \ \ \ \
\hat{H}=\left(
\begin{tabular}{c}
$\hat{H}_L$\\
$\hat{H}_R$
\end{tabular}
\right),
\end{equation}
where $H_{L,R}$ and
$\hat{H}_{L,R}$ are two component objects which are
charged under  ${\rm SU}(2)_L\times {\rm SU}(2)_R\times {\rm U}(1)_{B-L}$ as

\begin{equation}
    H_L {\rm \ and\ }\hat{H}_L: ({\bf 2},{\bf 1},1),\ \ \
    H_R {\rm \ and\ }\hat{H}_R: ({\bf 1},{\bf 2},1).
\end{equation}
$H$ couples to both the gauge boson sector and the fermion sector while $\hat{H}$ couples to the gauge boson sector only (for reasons discussed below).   This can be achieved by imposing a discrete symmetry under which $\hat{H}$ is odd while all other particles are even.
Each Higgs acquires a non-zero vacuum expectation value (vev) as  ($\hat{f}\gg f$)
\begin{eqnarray}\label{eq:vev1}
    <H> = \left(%
\begin{array}{c}
  0 \\
  0 \\
  0 \\
  f \\
\end{array}%
\right),\;\;\;\;\;
<\hat{H}> = \left(%
\begin{array}{c}
  0 \\
  0 \\
  0 \\
  \hat{f} \\
\end{array}%
\right),
\end{eqnarray}
breaking one of the U(4) to U(3), respectively.    Each yields seven
Nambu-Goldstone bosons and one massive radial mode. 
The Higgs vevs also break ${\rm SU}(2)_R\times {\rm U}(1)_{B-L}$ down to
the SM ${\rm U}(1)_Y$.

Below the cutoff scale $\Lambda$, the
radial modes are integrated out and the effective theory can be
described by a nonlinear sigma model of the 14 Goldstone bosons. 
We follow the parametrization of the scalar fields $H$ and $\hat{H}$ as described in 
Ref.~\cite{twinhiggssugoh}.  After spontaneous global symmetry breaking, three Goldstone bosons 
are eaten by the massive gauge bosons $W_H^\pm$ and $Z_H$,  and become their longitudinal components.
After electroweak symmetry breaking,  three additional Goldstone 
bosons are eaten by the SM gauge bosons $W^\pm$ and  $Z$.
With certain
re-parametrizations of the fields,  we are left with four Higgses that couple to both the fermion sector and the gauge boson sector: one neutral pseudoscalar 
$\phi^0$, a pair of charged scalars $\phi^\pm$, and the SM
physical Higgs $h_{SM}$.  In addition, an ${\rm SU}(2)_L$ doublet
$\hat{h}=(\hat{h}_1^+, \hat{h}_2^0)$, which resides in $\hat{H}_L$, is charged under an additional discrete parity and couples to the gauge boson sector only.

The left-right discrete
symmetry ensures that the global symmetry is respected for the one-loop
quadratic order of the Higgs potential, and so the quadratically divergent corrections to the
masses of the Goldstone bosons (i.e., the Higgses) are absent. The sub-leading contribution is only
proportional to $\ln\Lambda$, for $\Lambda$ being the cutoff
scale.  No severe fine tuning is introduced for $\Lambda$ of the
order of  5$-$10 TeV.

Note that in the LRTH model, it is necessary to introduce both the Higgs
fields $H$ and $\hat{H}$, with a hierarchy between their vevs: $\hat{f} \gg f$.  The reason is that the masses  of the heavy gauge bosons, which are proportional to $g\hat{f}$,  need be 
large enough to avoid the constraints from the electroweak
precision measurements.  Such a large value for $\hat{f}$ does not
reintroduce the fine tuning problem for the Higgs mass, since
the gauge boson contributions to the Higgs mass are suppressed
by the smallness of the gauge couplings. 
The Higgs
field $\hat{H}$, however, cannot  couple to the SM fermions, in particular, the top quark sector.  Otherwise, the heavy top quark (the fermionic top quark partner that is introduced in the LRTH model) 
obtains a much larger
mass of the order of $y\hat{f}$, which could result  in a large, dangerous contribution to the Higgs mass from the top sector.  
To avoid this, a parity is introduced in
the model under which $\hat{H}$ is odd while all other fields are
even. This parity thus forbids renormalizable couplings between
$\hat{H}$ and fermions, especially the top quark sector. Therefore,
$\hat{H}$ couples only to the gauge boson sector, while $H$
couples to both the gauge sector and the matter fields.  The ${\rm SU}(2)_L$ doublet $\hat{h}$, which resides in $\hat{H}_L$, is also odd under this additional matter parity.  The lightest
particle in $\hat{h}$,  typically one of the neutral components, is stable, and therefore constitutes  a good dark matter candidate.

\section{Dark matter candidate}
\label{sec:darkmatter}

The masses for $\hat{h}_1^\pm$ and $\hat{h}_2^0$, which are
relevant for the dark matter relic density analysis, can be
obtained from the one-loop CW potential arising from  gauge boson contributions. 
At leading order,  the Higgs masses are\cite{twinhiggssugoh}:
\begin{eqnarray}
{m}_{\hat{h}_2}^2 & \sim & \frac{3}{16\pi^2}\frac{m_{W_H}^2}{2}\left[
    g_2^2(\ln\frac{\Lambda^2}{m_{W_H}^2}+1)+\frac{2g_1^2+g_2^2}{2}(\ln\frac{\Lambda^2}{m_{Z_H}^2}+1)\right],
\label{eq:higgsmass1}\\
{m}_{\hat{h}_1}^2
&\sim&{m}_{\hat{h}_2}^2+\frac{3}{16\pi^2}
g_1^2m_W^2(\frac{m_{W_H}^2}{m_{Z_H}^2}
\ln\frac{m_{Z_H}^2}{m_{Z}^2}+\ln\frac{\Lambda^2}{m_{Z_H}^2}+1),
\label{eq:higgsmass2}
\end{eqnarray}
where $m_{W,W_H,Z,Z_H}$ are the gauge boson masses.   
The gauge couplings $g_1$ and $g_2$ are 
related to $e$ and Weinberg angle $\theta_w$ by
\begin{equation}
    g_1=\frac{e}{\sqrt{\cos2\theta_w}},\ \ \
    g_2=\frac{e}{\sin\theta_w}.
\label{eq:coupling}
\end{equation}
The cutoff scale $\Lambda$ is typically taken to be $4 \pi f$.
Numerically, the masses for 
$\hat{h}_1^\pm$ and $\hat{h}_2^0$ are around 200 $-$ 700 GeV for typical values of 
$m_{W_H}$, $m_{Z_H}$ in the model.
The small mass splitting between $\hat{h}_1^\pm$ and $\hat{h}_2^0$  [2nd term in Eq.~(\ref{eq:higgsmass2})]  is
around a few hundred MeV, which is due to electromagnetic interactions
that are absent for $\hat{h}_2^0$.

In addition, we can introduce a term in the Higgs potential
\begin{equation}\label{eq:muterm}
   V_H= \hat{\mu}^2\hat{H}_L^{\dagger}\hat{H}_L,
\end{equation}
which gives additional masses for $\hat{h}_1^\pm$ and $\hat{h}_2^0$ as 
\begin{eqnarray}
    \Delta m_{\hat{h}_1}^2 &\sim & \Delta m_{\hat{h}_2}^2 \sim \hat{\mu}^2.
\end{eqnarray}
This allows us to vary the mass of the dark matter 
independently as a free
parameter. However, such a term breaks the left-right symmetry softly.
Therefore, it is natural for it not to be much bigger than $f$ (typically between 500 GeV and 1500 GeV).
Note that since $ \hat{\mu}^2$ could have either sign, the total mass for $\hat{h}_1^\pm$ and $\hat{h}_2^0$ could be either larger or smaller than the contributions from the CW potential in Eqs.~(\ref{eq:higgsmass1}) and (\ref{eq:higgsmass2}).

\begin{figure}[bht]
\begin{center}
\resizebox{2.in}{!}{\includegraphics*[60,570][250, 740]{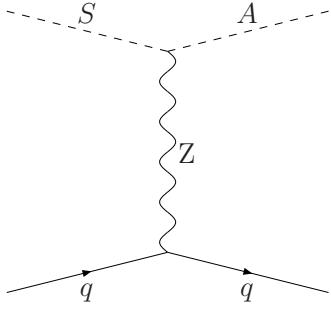}}
\caption{Feynman diagram that contributes to the elastic scattering of the dark matter with nuclei.}
\label{fig:SIscattering}
\end{center}
\end{figure}

The complex scalar $\hat{h}_2^0$ can be written as 
\begin{equation}
\hat{h}_2^0=\frac{\hat{S}+i\hat{A}}{\sqrt{2}}, 
\end{equation}
where $\hat{S}$ and $\hat{A}$ are the scalar and pseudoscalar fields, respectively. 
The cross section for scattering between $\hat{S}$ (or $\hat{A}$) with matter via $Z$ exchange (see Fig.~\ref{fig:SIscattering}) is about eight to nine  orders of magnitude larger than the limit  of spin independent interactions from current dark matter direct detection experiments\cite{SIresult}: $\sigma_{SI} \lesssim 10^{-42} {\rm cm}^2$ at 90\% C.L..  Such constraints, however, can be avoided if there is a small mass splitting between $\hat{S}$ and $\hat{A}$: 
$\delta_{2} = m_{\hat{A}}-m_{\hat{S}} \geq {\rm a\ few}$ hundred MeV, given the typical kinetic energy 
of the dark matter and the momentum transfer between the dark matter and the scattering nuclei in such scattering processes.  Here and in our numerical analysis below, we assume that $m_{\hat{S}} < m_{\hat{A}}$, therefore $\hat{S}$ is the dark matter candidate.  For $m_{\hat{S}} > m_{\hat{A}}$, the pseudoscalar $\hat{A}$ becomes the dark matter candidate.  However, the numerical results that we present for $\hat{S}$ also apply for the pseudoscalar dark matter.

\begin{figure}[bht]
\begin{center}
\resizebox{2.in}{!}{\includegraphics*[50,510][250, 660]{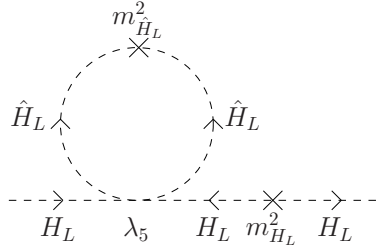}}
\caption{Contribution to the Higgs mass term $m_{H_L}^2 H_L H_L^\dagger$ from the
$\lambda_5$ quartic term in the Higgs potential. }
\label{fig:higgsmass}
\end{center}
\end{figure}

Such a mass splitting between $\hat{S}$ and $\hat{A}$, can be obtained by introducing a quartic term in the Higgs potential:  
\begin{equation}
V_H=-\frac{\lambda_5}{2} [(H_L^\dagger\hat{H}_L)^2+h.c.].
\end{equation}
The sign of the quartic term is picked such that $m_{\hat{A}}>m_{\hat{S}}$ for positive $\lambda_5$. 
Once $H_L$ obtains a vev $(0,v/\sqrt{2})$, it generates a splitting between $m_{\hat{S}}^2$ and $m_{\hat{A}}^2$:   
\begin{equation}
m_{\hat{A}}^2-m_{\hat{S}}^2 = \lambda_5 v^2.
\end{equation}
Introducing such a term in the Higgs potential breaks the left-right symmetry explicitly.  Therefore,  it could give a potentially dangerous quadratically divergent contribution to the Higgs mass.  
However, quartic terms $-(\lambda_5/4) h^2 \hat{S}^2$  and $(\lambda_5/4) h^2 \hat{A}^2$ have opposite signs, which ensures the  cancellation between the quadratically divergent contribution from the $\hat{S}$ loop and that from the $\hat{A}$ loop.  This  leaves only logarithmic contributions which are safe for $\lambda_5 \sim 1$.  Another way to understand this cancellation is that at one-loop level, a four point interaction 
($\lambda_5/2) (H_L^\dagger \hat{H}_L)^2$ (both $H_L$ fields flow out) cannot give rise to the
operator $m_{H_L}^2 H_L^\dagger H_L$ (with one $H_L$ flows in and another $H_L$ flows out) unless there is a mass insertion of  $m_{H_L}^2 H_L^\dagger H_L$ to flip the field flow (see Fig.~\ref{fig:higgsmass}).  
Dimensional analysis tells us that we can at most get $m_{H_L}^2\log\Lambda$ terms from such a diagram while the $\Lambda^2$ contribution is absent.  Given that $\lambda_5$ is almost not constrained, we can get a large splitting between $m_{\hat{S}}$ and $m_{\hat{A}}$ via such a $\lambda_5$ quartic interaction.

Due to the small mass splitting (of about a few hundred MeV) between $m_{\hat{h}_1}$ and  $m_{\hat{h}_2}$, introducing a mass splitting between $\hat{S}$ and $\hat{A}$ also leads to a mass splitting between the charged Higgses $\hat{h}_1^\pm$ and the dark matter candidate $\hat{S}$ (or $\hat{A}$, if $m_{\hat{S}} > m_{\hat{A}}$).  Letting $\delta_{1} = m_{\hat{h}_1}-m_{\hat{S}}$, we have $\delta_{2}\approx 2 \delta_{1}$ for $\delta_{2}\gtrsim 1$ GeV.  On the other hand, we could also introduce a quartic term:
\begin{equation}
V_H=\lambda_4 |\hat{H}_L^\dagger H_L|^2.
\end{equation}
If  $H_L$ obtains a vev $(0,v/\sqrt{2})$, it generates
a mass splitting between $m_{\hat{h}_1}^2$ and $m_{\hat{h}_2}^2$ (in addition to the mass splitting introduced by the ${\rm U}(1)_{em}$ in the CW potential): 
\begin{equation}
m_{\hat{h}_1}^2-m_{\hat{h}_2}^2 = -\frac{\lambda_4}{2} v^2. 
\end{equation}
In the LRTH model, 
such a term also breaks the left-right symmetry which could produce a dangerous contribution to the Higgs mass if $\lambda_4$ is too large. 
Unlike $\lambda_5$, which is not strongly constrained as discussed above, a simple dimensional analysis shows that $|\lambda_4| \leq (f/\Lambda)^2$.  For the cutoff scale $\Lambda$ taken to be $4 \pi f$, 
$|\lambda_4| \leq 1/(16 \pi^2)$.  Therefore, in the LRTH model, the corresponding mass splitting between $\hat{h}_1^\pm$ and $\hat{h}_2$ is at most  about a few GeV, typically much smaller than $\delta_2$.  Including $\lambda_4$ and $\lambda_5$ terms, and the ${\rm U}(1)_{em}$ contribution in the CW potential, we get 
\begin{equation}
m_{\hat{h}_1}^2 - m_{\hat{S}}^2 = \frac{\lambda_5}{2} v^2 - \frac{\lambda_4}{2} v^2 + 
{\rm small\ CW\ contribution}.
\end{equation}
In the LRTH model, when $\lambda_5$ could be large while $\lambda_4$ is constrained to be small, the approximate relation $\delta_{2} \approx  2 \delta_{1}$ holds for not so small values of $\delta_2$.  In the general IHDM, however, such an approximate relation could be violated.  In our analysis below, we study  both the most general case that $\delta_{1}$ and $\delta_{2}$ are two independent parameters, as in the general IHDM, as well as the case when the relation $\delta_{2} \approx  2 \delta_{1}$ holds, as 
in the LRTH model.

\section{Dark Matter Analysis}
\label{sec:analysis}

We analyzed the relic density using the program MicrOMEGAs\cite{micromegas}.  This program solves the Boltzmann equation numerically, using  the program CalcHEP\cite{calchep} to calculate all  the relevant cross sections.  The CalcHEP model files for the LRTH model can be found at 
Ref.~\cite{LRTHmodel}.  When the mass splittings between the dark matter candidate and other particles are small, coannihilation effects are also included.    We have identified two mass regions of $m_{\hat{S}}$ that could provide the amount of dark matter relic density that is consistent with the WMAP result at the 3 $\sigma$ level:  (A) low mass region where $m_{\hat{S}}<$ 100 GeV, and  (B) high mass region where 400 GeV $ < m_{\hat{S}} <$ a few TeV.

\subsection{Low Mass Region: $m_{\hat{S}}<$ 100 GeV}
 
\begin{figure}[bht]
\begin{center}
\resizebox{3.in}{!}{\includegraphics*[83,230][510,562]{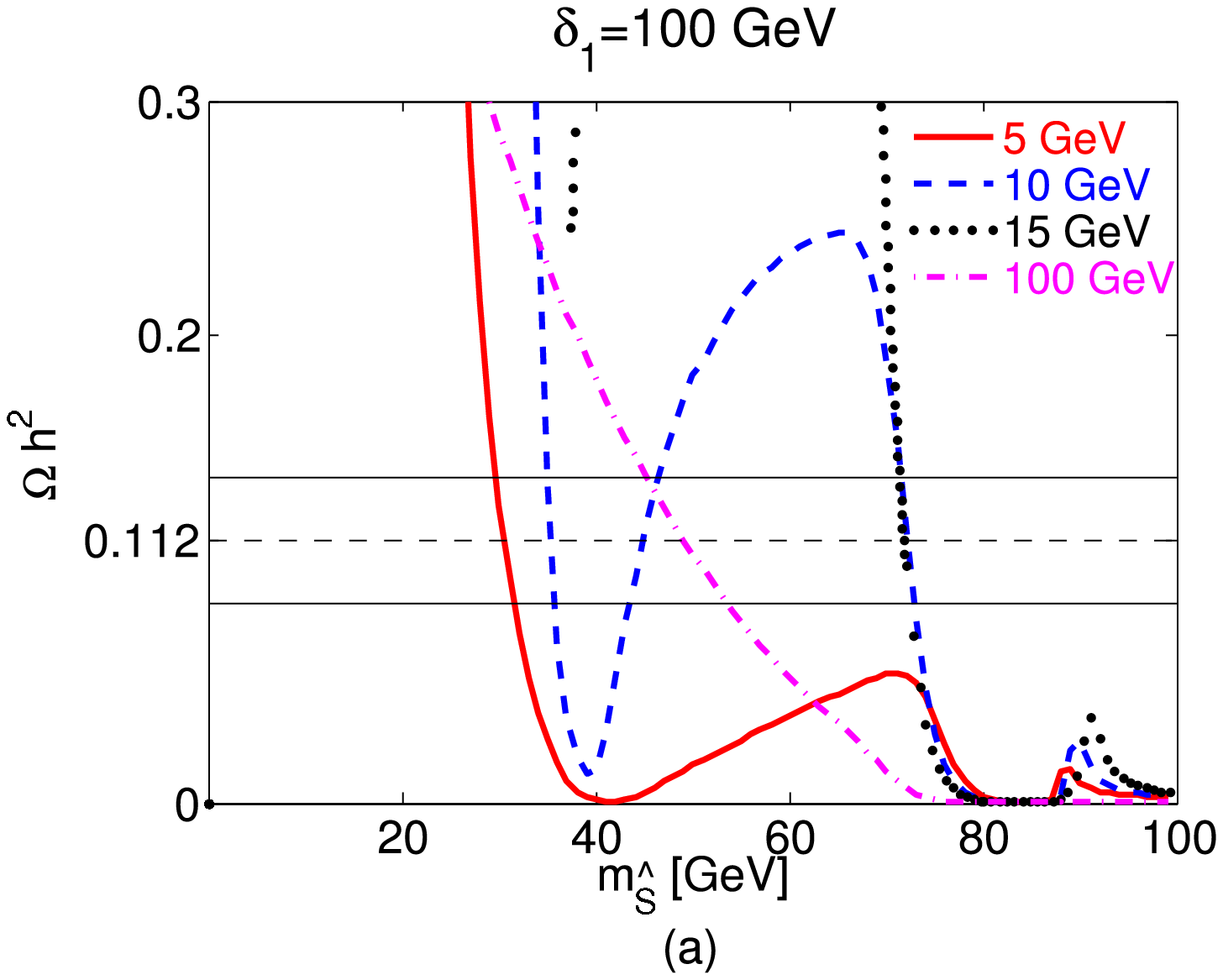}}
\resizebox{3.in}{!}{\includegraphics*[83,230][510,562]{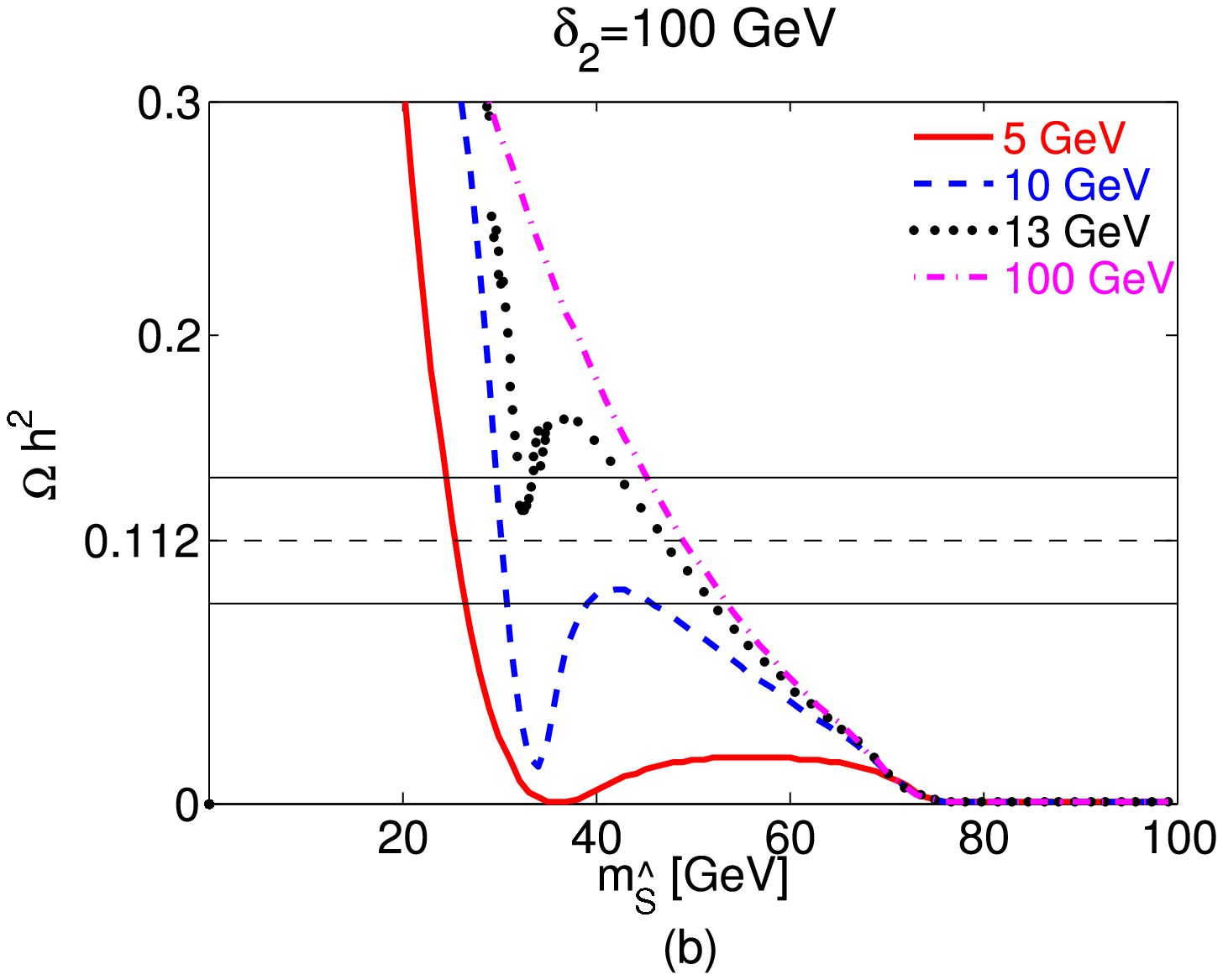}}
\caption{Plot of the relic density as a function of $m_{\hat{S}}$ for the low mass region.  The left plot (a)  fixes $\delta_1$ at 100 GeV while $\delta_2$ is set to be 5 GeV (solid), 10 GeV (dashed), 15 GeV (dotted), and 100 GeV (dash-dotted). 
The right plot (b) fixes $\delta_2$ at 100 GeV while $\delta_1$ is set to be 5 GeV (solid), 10 GeV (dashed), 13 GeV (dotted), and 100 GeV (dash-dotted).  The horizontal band shows the 3 $\sigma$ region of the WMAP result: 
$\Omega h^2=0.112 \pm 0.027$. 
}
\label{fig:reliclow}
\end{center}
\end{figure}

Let's first study the most general case when $\delta_1$ and $\delta_2$ are two independent parameters. 
In Fig.~\ref{fig:reliclow}, we show the relic density dependence on the mass of the dark matter $m_{\hat{S}}$ for (a) $\delta_1$=100 GeV (when coannihilation  between $\hat{S}$ and $\hat{h}_1^\pm$ can be ignored), and various values for $\delta_2$, and
(b) $\delta_2$=100 GeV (when coannihilation  between $\hat{S}$ and $\hat{A}$ can be ignored), and various   values for $\delta_1$.


\begin{figure}[hbt]
\begin{center}
\resizebox{4.in}{!}{\includegraphics*[71,520][554,706]{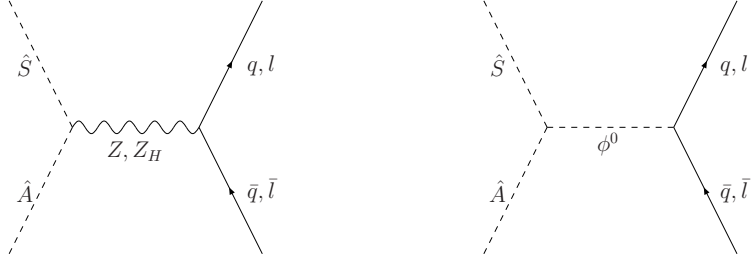}}
\caption{Feynman diagrams that contribute to the coannihilation processes: $\hat{S}\hat{A}\rightarrow q\bar{q}/l\bar{l}$.  }
\label{fig:coann}
\end{center}
\end{figure}

For $\delta_1=100$ GeV and $\delta_2=10$ GeV [dashed curve in Fig.~\ref{fig:reliclow}~(a)],  coannihilation between $\hat{S}$ and $\hat{A}$ needs to be taken into account due to the relatively small mass splitting $\delta_2$. 
There are three mass windows where $\Omega h^2$ falls within the WMAP bounds.   
The first two mass windows  near $m_{\hat{S}}\sim 40$ GeV receive dominant contributions from the coannihilation of $\hat{S}\hat{A} \rightarrow q\bar{q}/l\bar{l}$ via $Z$ exchange, as shown in Fig.~\ref{fig:coann}.   The contribution from $Z_H$ exchange is suppressed by the heavy $Z_H$ mass, while the contribution from $\phi^0$ exchange is suppressed due to the small $\hat{S}\hat{A}\phi^0$ coupling.   
The dip between these two mass windows is due to the Z pole: $m_{\hat{S}}+m_{\hat{A}} = m_Z$.
We denote these mass windows as ``low mass $Z$ pole region".

\begin{figure}
\begin{center}
\resizebox{4.in}{!}{\includegraphics*[60,500][556,650]{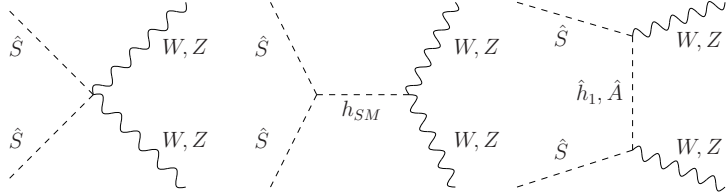}}
\caption{Feynman diagrams that contribute to the annihilation processes 
$\hat{S}\hat{S}\rightarrow{WW/ZZ}$.  }
\label{fig:annSSWW}
\end{center}
\end{figure}

Once $m_{\hat{S}}$ increases above the $Z$ pole region, the coannihilation cross section decreases, leading to a relic density $\Omega h^2$ above the WMAP region.  However, when $m_{\hat{S}}$ is
close to the $W$ boson mass, a new annihilation channel  $\hat{S}\hat{S} \rightarrow WW$ opens, which brings the relic density down to the WMAP region [the third mass window: $m_{\hat{S}}\sim 75$ GeV,  of the dashed curve in Fig.~\ref{fig:reliclow}~(a)].  We denote this mass window as ``low mass $WW$ bulk region".  The corresponding Feynman diagrams are shown in  Fig.~\ref{fig:annSSWW}.  When $m_{\hat{S}}$ further increases, channel  $\hat{S}\hat{S} \rightarrow ZZ$ also becomes accessible, which further increases the annihilation cross section, reducing the relic density.  
When $2m_{\hat{S}}=m_{h_{SM}} \sim 170 $ GeV (as in the LRTH model), the dominant contribution comes from the exchange with 
an $h_{SM}$ pole and the relic density reaches its minimum.   The slight rise in $\Omega h^2$ around 90 GeV is due to the destructive interference between the $\hat{S}\hat{S} \rightarrow WW/ZZ$ Feynman diagrams.

\begin{figure}
\begin{center}
\resizebox{2.in}{!}{\includegraphics*[67,520][300,700]{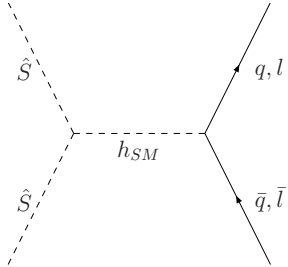}}
\caption{Feynman diagrams that contribute to the annihilation processes 
$\hat{S}\hat{S}\rightarrow{q\bar{q}/l\bar{l}}$.}
\label{fig:annSSqq}
\end{center}
\end{figure}

For smaller value of  $\delta_2$, coannihilation effects become more important.  
The solid curve in Fig.~\ref{fig:reliclow}~(a) shows the $\Omega h^2$ dependence on $m_{\hat{S}}$ for
$\delta_2=5$ GeV.  The contribution from the coannihilation cross section becomes so large that only part of the low mass $Z$ pole region survives.  

As $\delta_2$ is increased, coannihilations becomes less effective.   The dotted curve in Fig.~\ref{fig:reliclow}~(a) shows the relic density for $\delta_2=15$ GeV, when coannihilations can be ignored.   The low mass $Z$ pole region no longer exists for such  large values of $\delta_2$.  The relevant annihilation process for the low mass region (before annihilation into $WW/ZZ$ opens up) is $\hat{S}\hat{S}\rightarrow q\bar{q}/l\bar{l}$ via $h_{SM}$ exchange (Fig.~\ref{fig:annSSqq}).  The cross section is small due to the suppressed coupling of $\hat{S}\hat{S}h_{SM}$ and the small Yukawa  couplings of $h_{SM}q\bar{q}$ and $h_{SM}l\bar{l}$, which leads to a relic density larger than the WMAP preferred value.   The low mass $WW$ bulk region, however,  still remains once $m_{\hat{S}}$ is close to the $W$ boson mass.

As  $\delta_2$ further increases, process $\hat{S}\hat{S} \rightarrow q\bar{q}/l\bar{l}$ becomes more and more important due to the enhanced coupling $\hat{S}\hat{S}h_{SM}$, which is proportional to $\delta_2$.    The dash-dotted curve in Fig.~\ref{fig:reliclow}~(a) shows the relic density for 
$\delta_2=100$ GeV,  when the relic density falls into the WMAP window for a smaller value of $m_{\hat{S}}$: $m_{\hat{S}}\sim$ 50 GeV.  The dominating process for this region is $\hat{S}\hat{S} \rightarrow b\bar{b}$.  We denote this mass window as ``low mass $bb$ bulk region", distinguishing from the  low mass $WW$ bulk region, which appears for smaller values of $\delta_2$. 
Typically,  for  $\delta_2\lesssim$ 50 GeV, $\hat{S}\hat{S} \rightarrow WW$  is the dominating annihilation process, which prefers a larger $m_{\hat{S}}$ around 70 GeV for the WMAP window.   For  $\delta_2\gtrsim$50 GeV,    $\hat{S}\hat{S} \rightarrow b\bar{b}$ starts to dominate, when the WMAP mass window shifts to smaller values of $m_{\hat{S}}$.    As we will show later, the switching between these two regions occurs at lower $\delta_2$ once $\delta_1$ becomes smaller.  There is, however, no clear separation of the low mass $bb$ bulk region and the low mass $WW$ region since in general both processes contribute to dark matter annihilations.

Part of the small $m_{\hat{S}}$ region, however,  has already been excluded when we impose Z width constraints.  For $m_{\hat{S}}+m_{\hat{A}}<m_Z$, 
process $Z \rightarrow \hat{S}+\hat{A}$ contributes to the $Z$ decay width $\Gamma_Z$.   On the other hand, $\Gamma_Z$ has been measured very precisely at the LEP\cite{pdg}, which agrees very well with the SM prediction.  Therefore, there is little room for any new $Z$ decay process.  The region that corresponds to $2 m_{\hat{S}}+\delta_2<m_Z$ is excluded once we impose the $Z$ width constraints.  Therefore, part of the low mass $Z$ pole region is ruled out by the $Z$ decay width consideration. 
 
Fig.~\ref{fig:reliclow}~(b) shows the relic density as a function of $m_{\hat{S}}$ for various splittings of $\delta_1$, with $\delta_2$ fixed at 100 GeV.   
The coannihilation effects between $\hat{S}$ and $\hat{A}$ can now be safely ignored, while coannihilation effects between $\hat{S}$ and $\hat{h}_1^\pm$ have to be taken into account when $\delta_1$ is small enough.  We obtain results similar to those presented in Fig.~\ref{fig:reliclow}~(a).  The low mass pole region appears for a lower $m_{\hat{S}}$ region since it corresponds to coannihilation process $\hat{S}\hat{h}_1^\pm\rightarrow
q \bar{q}^\prime$ via $W$ exchange instead of $Z$ exchange.   We indicate this as ``low mass $W$ pole region".   For larger values of $\delta_1\gtrsim$ 15 GeV,  the coannihilation between $\hat{S}$ and $\hat{h}_1^\pm$ can be neglected, and the low mass $W$ pole region disappears.

Due to the large value of $\delta_2$, the low mass $bb$ bulk region  appears for $m_{\hat{S}} \sim $ 50 GeV. 
The low mass $WW$ bulk region does not appear for  $\delta_2=100$ GeV since $\hat{S}\hat{S}\rightarrow b\bar{b}$ already saturates the annihilation cross section at smaller $m_{\hat{S}}$.

The charged Higgs has been searched for at the LEP and the
Tevatron \cite{chargedhiggslimitlep, chargedhiggslimitcdf}.  
A lower mass bound of 74 $-$ 79 GeV at 95\% C.L. is obtained at the LEP \cite{chargedhiggslimitlep} considering $H^+\rightarrow c \bar{s}$, $H^+ \rightarrow \tau^+\nu$,  or $H^+ \rightarrow W^*A$.  
A more recent search at CDF\cite{chargedhiggslimitcdf} studied the charged Higgs produced in the top quark decay $t \rightarrow H^+ b$, with $H^+$ further decaying into a pair of quarks, leptons, or $W^+\phi$.   All of those searches, however, rely on the couplings of the charged Higgs with fermions, which is absent for 
$\hat{h}_1^\pm$ in both the LRTH model and the IHDM.  Therefore, the experimental limit on the charged Higgs mass does not apply.  On the other hand, $Z\rightarrow 
\hat{h}_1^+\hat{h}_1^-$ is allowed once the charged Higgs mass is below $m_Z/2$, which leads to the deviation of $Z$ decay width $\Gamma_Z$  from the SM predictions.  Therefore, $m_{\hat{h}_1}=m_{\hat{S}}+\delta_1 < m_Z/2$ is excluded by the consideration of $\Gamma_Z$\cite{pdg},  and part of the low mass $W$ pole region in Fig.~\ref{fig:reliclow}~(b) is not allowed given this constraint.

\begin{figure}[bht]
\begin{center}
\resizebox{3.in}{!}{\includegraphics*[83,230][510,562]{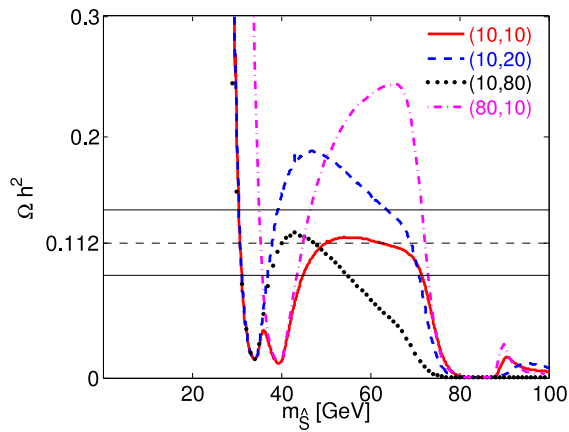}}
\caption{Plot of the relic density as a function of $m_{\hat{S}}$ for the low mass region, with $\delta_1$ and $\delta_2$ chosen as: $(\delta_1, \delta_2)=$ (10 GeV, 10 GeV) (solid curve), (10 GeV, 20 GeV)  (dashed curve), (10 GeV, 80 GeV) (dotted curve), and (80 GeV, 10 GeV) (dash-dotted curve).   The horizontal band shows the WMAP 3 $\sigma$ region.  }
\label{fig:reliclowdeltasmall}
\end{center}
\end{figure}

Fig.~\ref{fig:reliclowdeltasmall} shows the dependence of $\Omega h^2$ as a function of $m_{\hat{S}}$ for four different sets of $(\delta_1, \delta_2)$.   When $\delta_1$ and $\delta_2$ are both small, $(\delta_1, \delta_2)=$ (10 GeV, 10 GeV), as indicated by the solid curve, the coannihilations between $\hat{S}$, $\hat{A}$ and $\hat{h}_1^\pm$ should all be taken into account.  Therefore, there are two dips for $m_{\hat{S}}$ around 35 $-$ 40 GeV, which correspond to the $W$ pole and the $Z$ pole, respectively.  There is a relatively wide region of $m_{\hat{S}}$ that falls into the WMAP 3 $\sigma$ region for $50\ {\rm GeV} \lesssim m_{\hat{S}} \lesssim 70$ GeV, which is a combination of the pole region and the $WW$ bulk region.  For $(\delta_1, \delta_2)=$ (10 GeV, 20 GeV) (dashed curve), only the low mass $W$ pole region appears since $\delta_2$ is large enough that the coannihilation between $\hat{S}$ and $\hat{A}$ is ineffective.  However, $\delta_2$ is still relatively small such that the bulk region is dominantly $\hat{S}\hat{S}\rightarrow{W{W}}$.  When $\delta_2$ is set to be 80 GeV (dotted line), 
the bulk region is dominantly $\hat{S}\hat{S}\rightarrow{b\bar{b}}$ due to the relatively large value of $\delta_2$.  For $(\delta_1, \delta_2)=$ (80 GeV,10 GeV) (dash-dotted curve), only the low mass $Z$ pole region (due to the coannihilation of  $\hat{S}$ and $\hat{A}$) and low mass $WW$ bulk region appear.

\begin{figure}[bht]
\begin{center}
\resizebox{3.in}{!}{\includegraphics*[87,230][510,562]{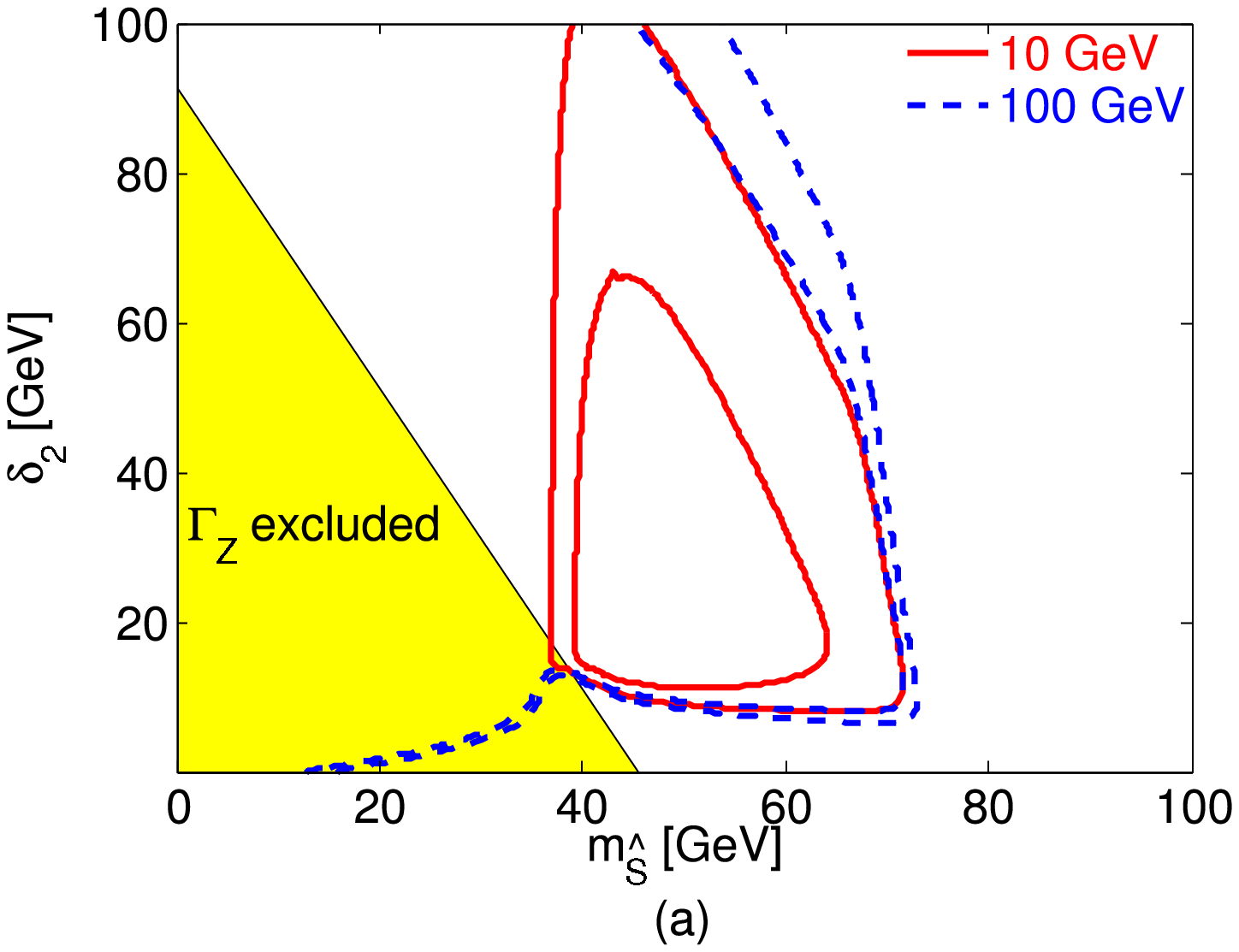}}
\resizebox{3.in}{!}{\includegraphics*[87,230][510,562]{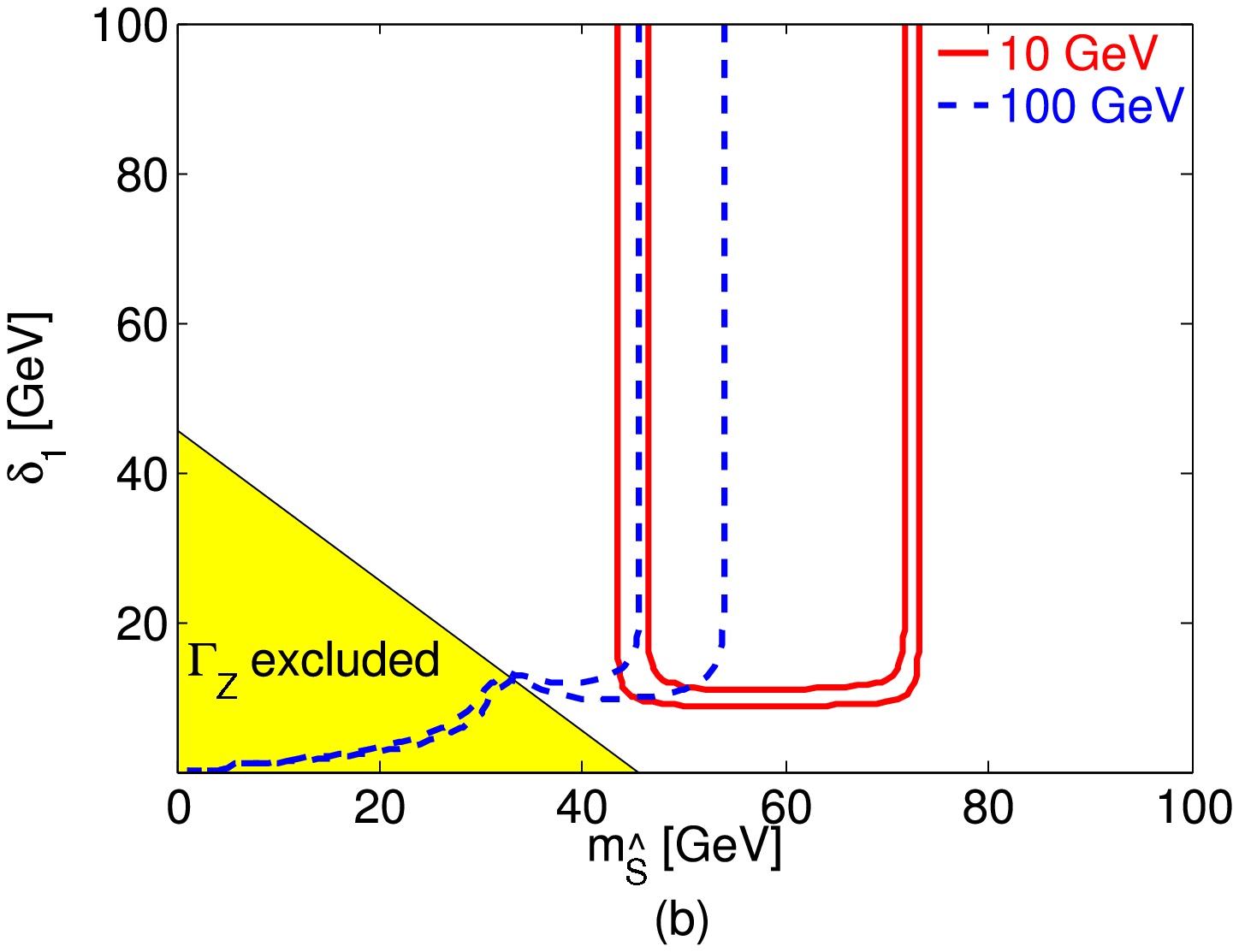}}
\caption{Contour plots of the relic density in the $m_{\hat{S}}$ vs. $\delta_2$ plane [plot (a)] and $m_{\hat{S}}$ vs. $\delta_1$ plane [plot (b)]  for the low mass region.  $\delta_1$ ($\delta_2$) is chosen to be 10 GeV (solid lines) and 100 GeV (dashed lines) for plot (a) [plot (b)]. The shaded region is excluded by experimental constraints from  $\Gamma_Z$.  The region enclosed by two contour lines corresponds to the WMAP 3 $\sigma$ region.   }
\label{fig:reliccontourlow}
\end{center}
\end{figure} 

Fig.~\ref{fig:reliccontourlow}~(a) shows the contour plot of $\Omega h^2$ in the $m_{\hat{S}}$ vs. $\delta_2$ plane for two values of $\delta_1$.   The region enclosed by two contour lines corresponds to the  WMAP 3 $\sigma$  region for the dark matter relic density.
The shaded region is excluded by the LEP constraints on the $Z$ decay width (We also applied $\Gamma_Z$ constraints for a given $\delta_1$.),  which eliminates part of the low mass $W$ or $Z$ pole region for small  $m_{\hat{S}}$.   

For $\delta_1=100$ GeV (regions enclosed by the dashed contour lines), the coannihilation between $\hat{S}$ and $\hat{h}_1^\pm$ can be ignored.    
For 7 GeV$\lesssim\delta_2\lesssim$13 GeV, there are two possible mass windows of  $m_{\hat{S}}$ for a given $\delta_2$, corresponding to the low mass $Z$ pole region and $WW$ bulk region as discussed earlier for Fig.~\ref{fig:reliclow}~(a).   For $\delta_2\gtrsim$13 GeV, there is only one mass window of $m_{\hat{S}}$ for a given $\delta_2$.  This corresponds to the case when the coannihilation effect is negligible and the dominating contribution comes from annihilation between dark matter themselves.  For $\delta_2\lesssim 50$ GeV, annihilation into $WW$ dominates, which leads to $m_{\hat{S}}$ around 70 GeV for the WMAP preferred region.  For larger $\delta_2$, annihilation into $b\bar{b}$ final states starts to contribute.  The curve bends to the left, leading to a WMAP mass window at a smaller value of $m_{\hat{S}}$.   
 
For $\delta_1$=10 GeV [solid curves in Fig.~\ref{fig:reliccontourlow}~(a)],  the vertical region around $m_{\hat{S}}=$ 40 GeV corresponds to the the low mass $W$ pole region. 
The bulk region shifted to the left comparing to $\delta_1=$ 100 GeV case.  For small $\delta_2$ around 10 GeV, the coannihilations between $\hat{S}$, $\hat{A}$, and $\hat{h}_1^\pm$ are effective.  There is a relatively large window of $m_{\hat{S}}$ between 40 to 75 GeV that could give the right amount of relic density.

Fig.~\ref{fig:reliccontourlow}~(b) shows the contour plot of $\Omega h^2$ in the $m_{\hat{S}}$ vs. $\delta_1$ plane for $\delta_2=$100 GeV (dashed curves) and 10 GeV (solid curves).  For larger values of $\delta_1\gtrsim 15$ GeV, the coanihilation between $\hat{S}$ and $\hat{h}_1^\pm$ can be ignored.   The $m_{\hat{S}}$ window is independent of $\delta_1$ since either the $Z$ pole region (for smaller $\delta_2$) or bulk region ($WW$ or $bb$) is independent of $\delta_1$.  For large $\delta_2=$100 GeV (region enclosed by the dashed curves), annihilation of $\hat{S}\hat{S}\rightarrow{b\bar{b}}$ results in a bulk mass window of $m_{\hat{S}}\sim 50$ GeV.  When $\delta_2$ gets smaller, the bulk mass window shifts to larger $m_{\hat{S}}$ as $\hat{S}\hat{S}\rightarrow{WW}$ becomes more and more important.  For $\delta_2=10$ GeV (regions enclosed by the solid curves), coannihilation of $\hat{S}$ and $\hat{A}$ needs to be taken into account, which leads to the appearance of the low mass $Z$ pole region for $m_{\hat{S}}$ around 45 GeV.   

\begin{figure}[bht]
\begin{center}
\resizebox{3. in}{!}{\includegraphics*[87,230][510,562]{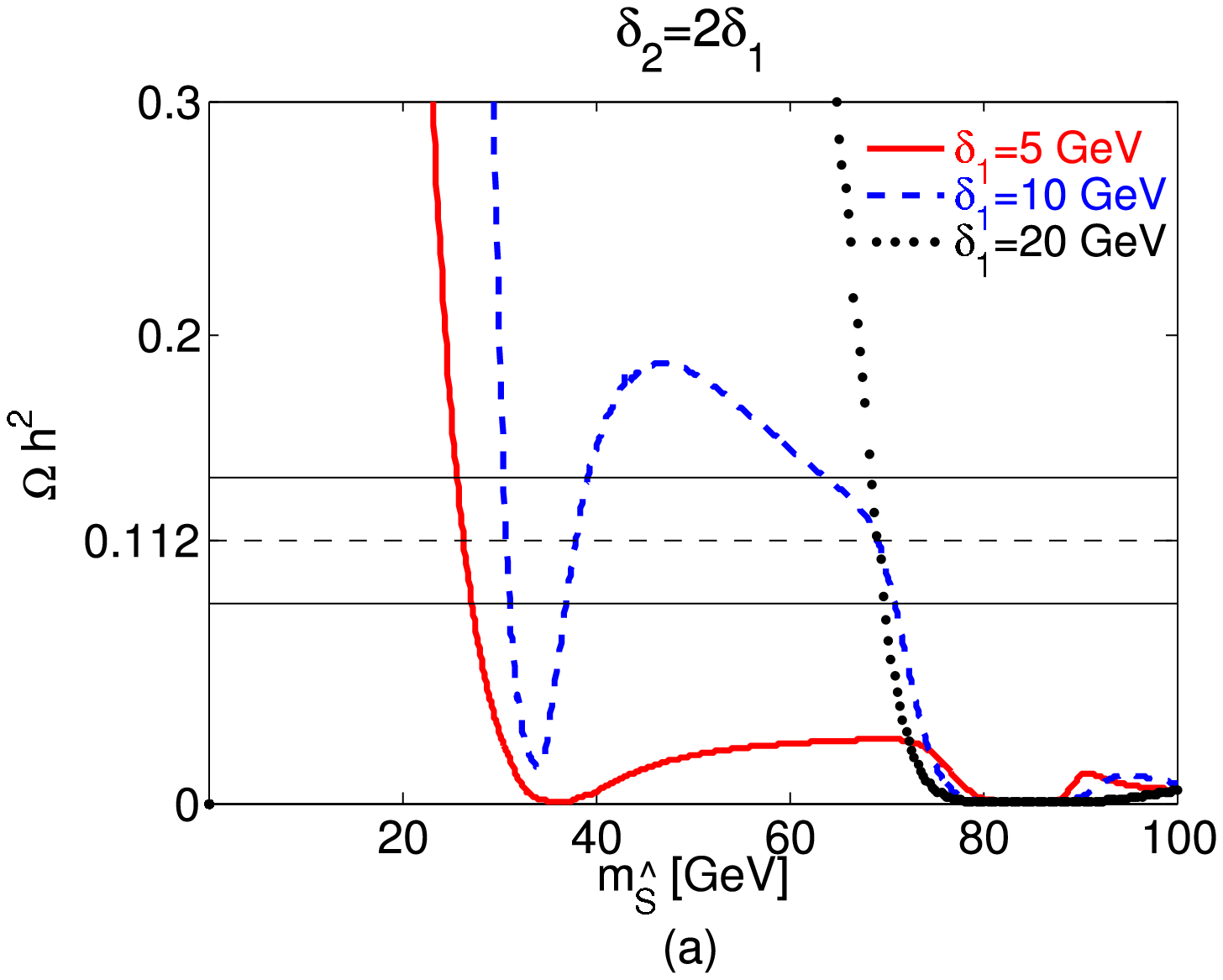}}
\resizebox{3.in}{!}{\includegraphics*[87,230][510,562]{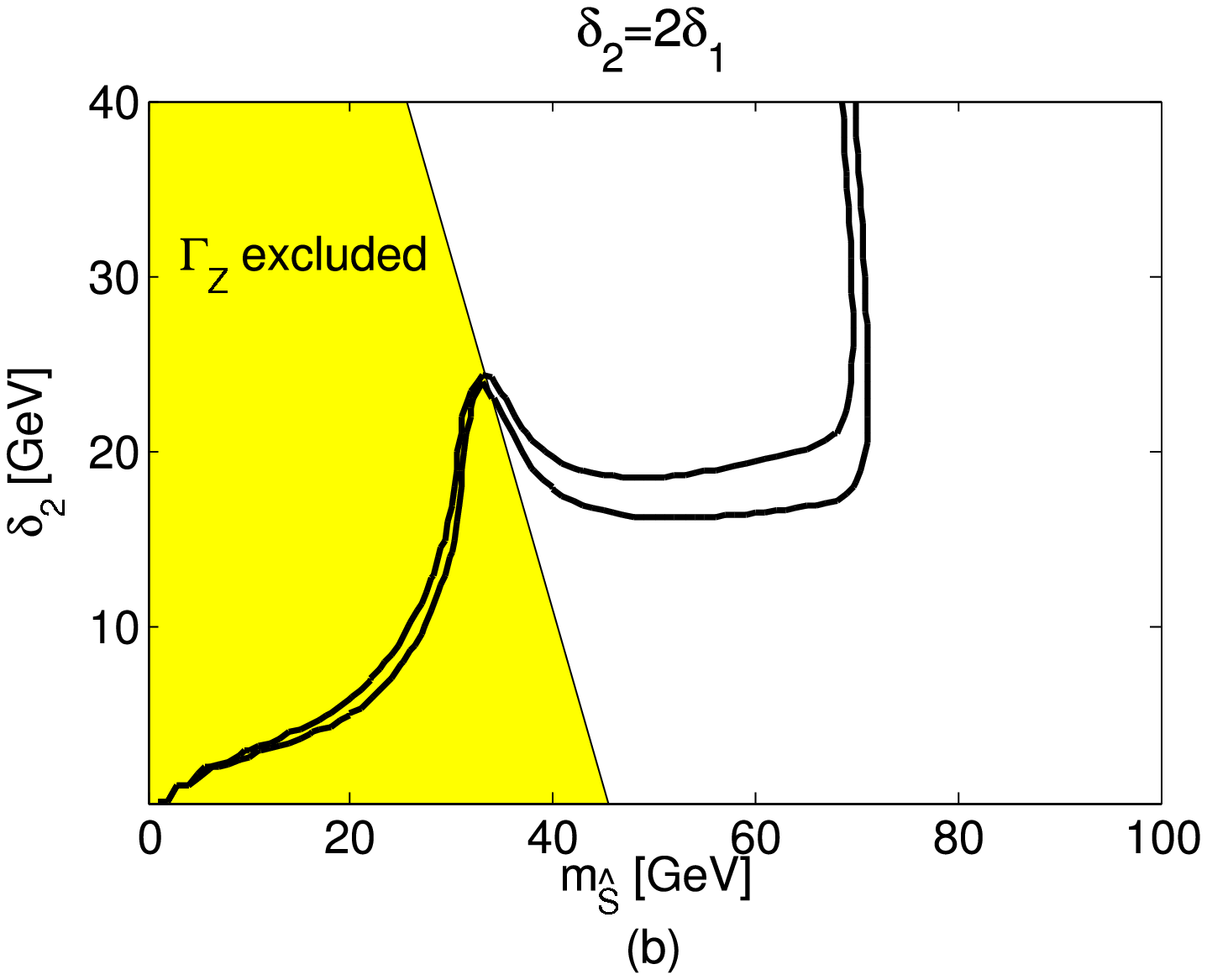}}
\caption{$\Omega h^2$ vs. $m_{\hat{S}}$ plot (a) and relic density contour plot (b) for $\delta_2 = 2 \delta_1$.  $\delta_1$ is taken to be 5 GeV(solid), 10 GeV(dashed), and 20 GeV(dotted) in plot (a).  The band in plot (a) and region enclosed by two contour lines in plot (b) are the  WMAP 3 $\sigma$ regions.}
\label{fig:relicLRTH}
\end{center}
\end{figure} 

In the LRTH model,  the mass splittings 
$\delta_{1,2}$ follow the approximate relation $\delta_2 = 2 \delta_1$ for not so small values of $\delta_2$.   The corresponding relic density plots are shown in Fig.~\ref{fig:relicLRTH}~(a) for the $m_{\hat{S}}$ dependence. 
For small $\delta_{1,2}$, the coannihilation between $\hat{S}$, $\hat{A}$, and $\hat{h}_1^\pm$ leads to the low mass pole region.  The low mass bulk region is  due to 
$\hat{S}\hat{S}\rightarrow {WW}$.  The corresponding WMAP 3 $\sigma$ region contour plot of the dark matter relic density is shown in Fig.~\ref{fig:relicLRTH}~(b).

\subsection{High Mass Region: 400 GeV$<m_{\hat{S}} <$ 1 TeV }

\begin{figure}[bht]
\begin{center}
\resizebox{3.1in}{!}{\includegraphics*[87,230][510,562]{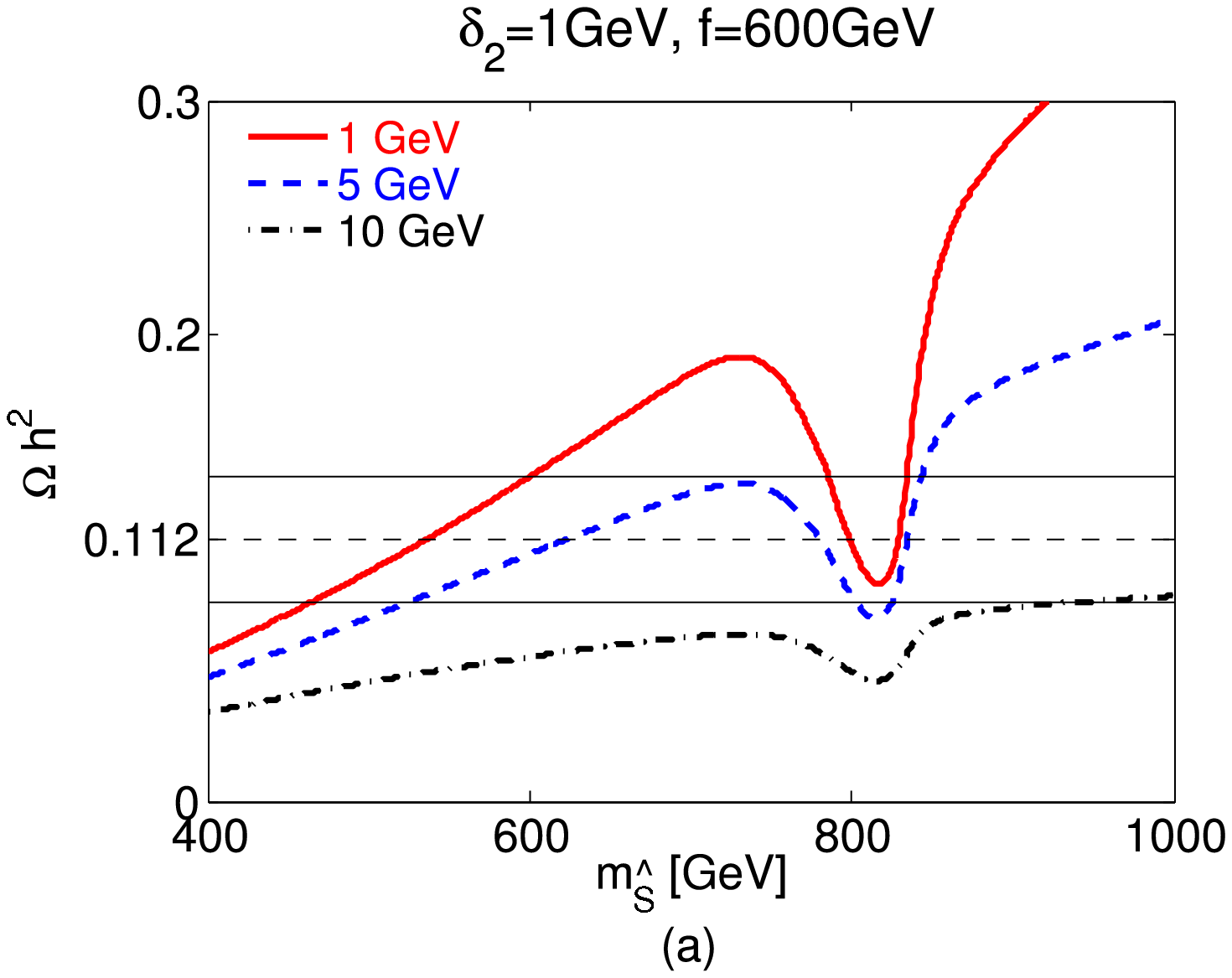}}
\resizebox{3.in}{!}{\includegraphics*[87,230][510,562]{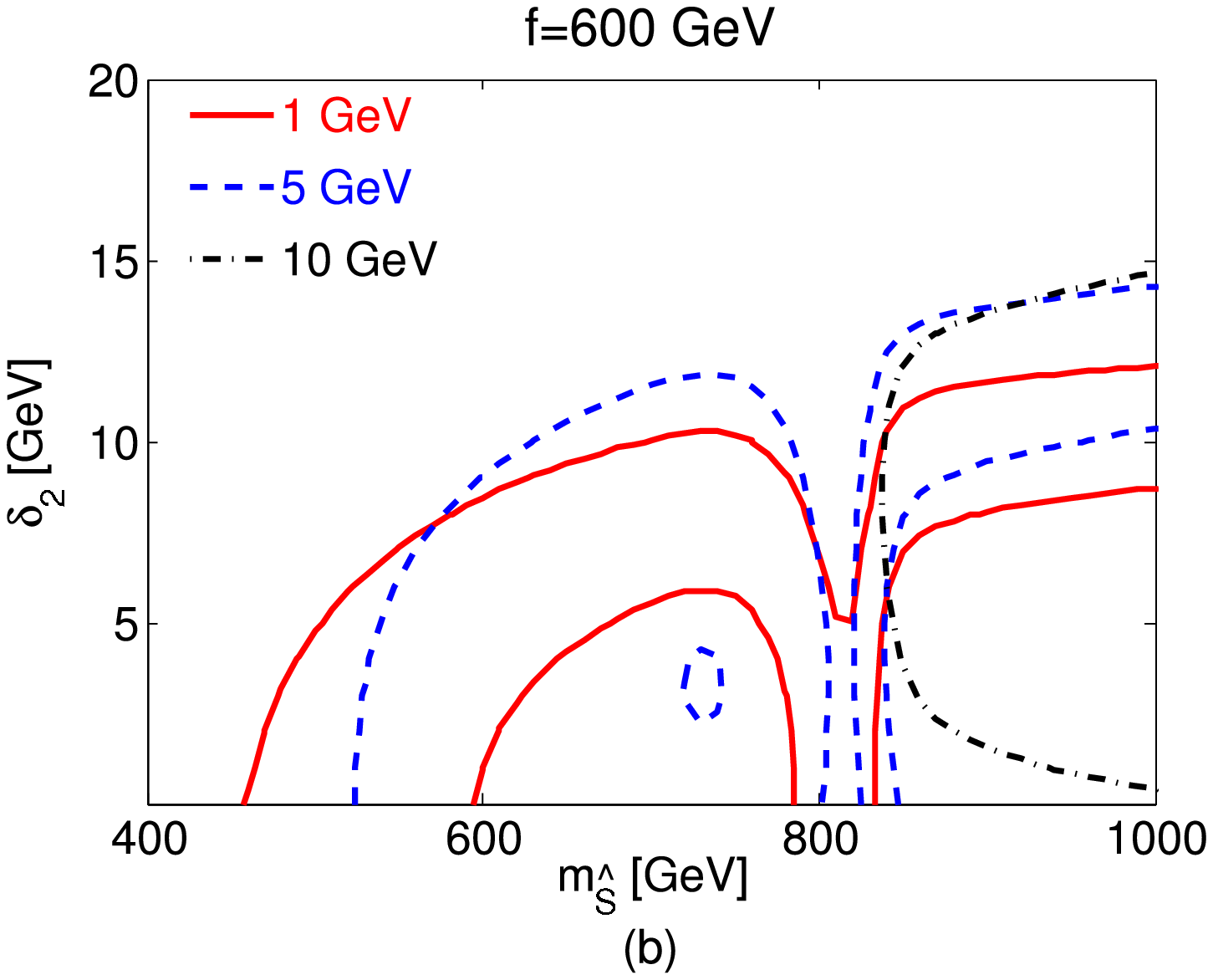}}
\caption{ Plot (a) shows the relic density $\Omega h^2$ vs. $m_{\hat{S}}$ for $\delta_2=1$ GeV.  Plot (b) shows the contour plot of the relic density in the $m_{\hat{S}}$ vs. $\delta_2$ plane.  $\delta_1$ is chosen to be 1  GeV (solid curve), 5 GeV (dashed curve),  and 10 GeV (dash-dotted curve) for both plots.  $f$ is chosen to be 600 GeV. 
The band in plot (a) and region enclosed by two contour lines in plot (b) are the  WMAP 3 $\sigma$ regions.}
\label{fig:relichigh1}
\end{center}
\end{figure} 

When the dark matter mass $m_{\hat{S}} > 400$ GeV, there are regions of parameter space that could also provide the right amount of dark matter relic density.  Fig.~\ref{fig:relichigh1}~(a) shows the relic density dependence on $m_{\hat{S}}$ for various  values of $\delta_1$.   There are typically two $m_{\hat{S}}$ mass windows. A ``high mass bulk region I" appears near $450\ {\rm GeV} < m_{\hat{S}} < 750\ {\rm GeV}$, with the dominating contributions coming from $\hat{S}\hat{S}\rightarrow{WW/ZZ}$.   A ``high mass pole region" appears for $m_{\hat{S}}$ around 810 GeV, where the coannihilation of $\hat{S}$ and $\hat{A}$ via the exchange of a heavy gauge boson $Z_H$ dominates (see Fig.~\ref{fig:coann} for the corresponding Feynman diagram).    When $\delta_1$ gets larger, the $\hat{S}\hat{S}$ annihilation cross section grows, which leads to a suppressed $\Omega h^2$ below the WMAP window.  However, at larger $m_{\hat{S}}\gtrsim$ 850 GeV [see dash-dotted curve in Fig.~\ref{fig:relichigh1}~(a)], a new bulk mass window, denoted as ``high mass bulk region II", might appear.
The numerical results of the dark matter relic density for fixed $\delta_1$ and varying $\delta_2$  are very similar to those presented in Fig.~\ref{fig:relichigh1}~(a).  The relic density decreases for increasing $\delta_2$ as well.    

Fig.~\ref{fig:relichigh1}~(b) shows the WMAP 3 $\sigma$ window in the $m_{\hat{S}}$ vs. $\delta_2$ plane for various values of $\delta_1$.  The pole region around $m_{\hat{S}} \sim 800$ GeV is independent of $\delta_{1,2}$.  For small values of $\delta_{1,2}$, the high mass bulk region I  always appears.  The high mass bulk region II, $m_{\hat{S}}\gtrsim 850$ GeV, on the other hand, only appears when either $\delta_1$ or $\delta_2$ is around 7$-$15 GeV.  For 
$\delta_{1,2}\gtrsim 15$ GeV, the annihilation cross sections become so big that the relic density always falls below the WMAP preferred region.

Note that the pole region only appears in the LRTH model, where the existence of the $Z_H$  pole dominates the coannihilation $\hat{S}\hat{A}\rightarrow q \bar{q}/l\bar{l}$ at $m_{\hat{S}}\sim m_{Z_H}/2$.  In the general IHDM, the pole region disappears.  However, the general results that we obtained above for the high mass bulk regions still apply. 

\begin{figure}[bht]
\begin{center}
\resizebox{3.in}{!}{\includegraphics*[83,230][510,562]{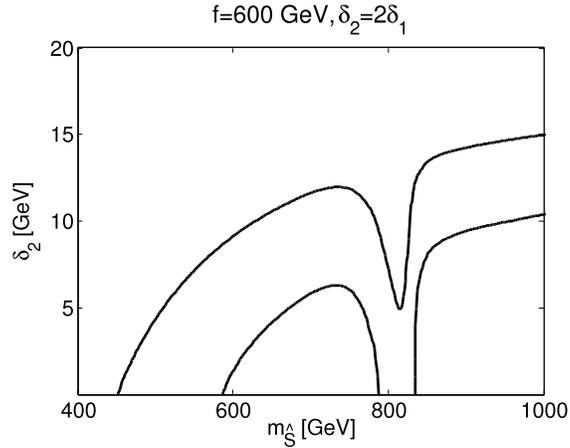}}
\caption{Contour plot of the relic density in the $m_{\hat{S}}$ vs. $\delta_2$ plane. LRTH relation of $\delta_2=2 \delta_1$ is imposed. The region enclosed by two contour lines  corresponds to the WMAP   3 $\sigma$  region. }
\label{fig:relichighLRTH}
\end{center}
\end{figure} 

Fig.~\ref{fig:relichighLRTH} shows the WMAP 3 $\sigma$ region in the $m_{\hat{S}}$ vs. $\delta_2$ plane when we impose the LRTH relation $\delta_2=2\delta_1$.  A relatively large region of parameter space could provide the right amount of dark matter relic density.

\begin{figure}[bht]
\begin{center}
\resizebox{3.in}{!}{\includegraphics*[87,230][510,562]{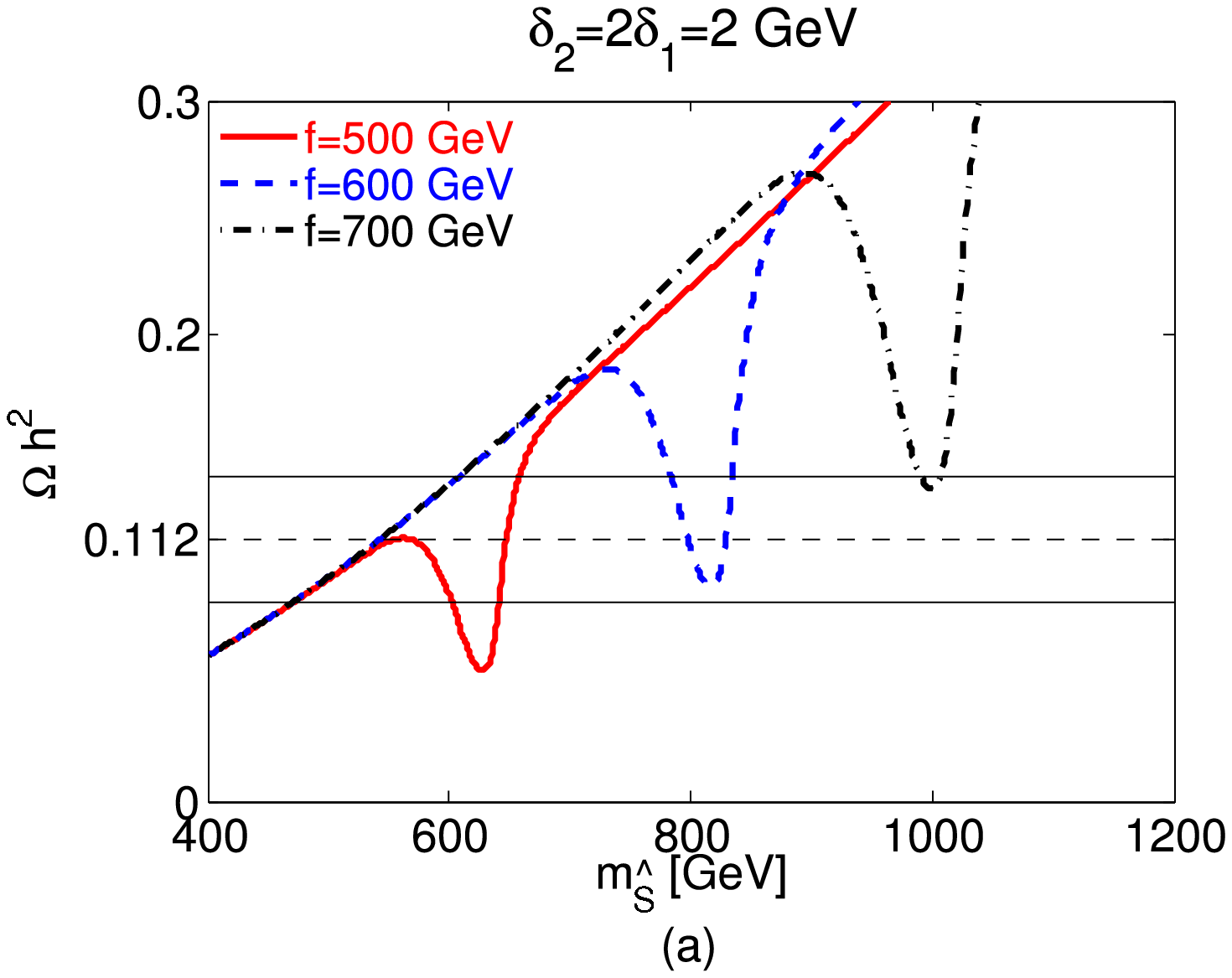}}
\resizebox{3.in}{!}{\includegraphics*[87,230][510,562]{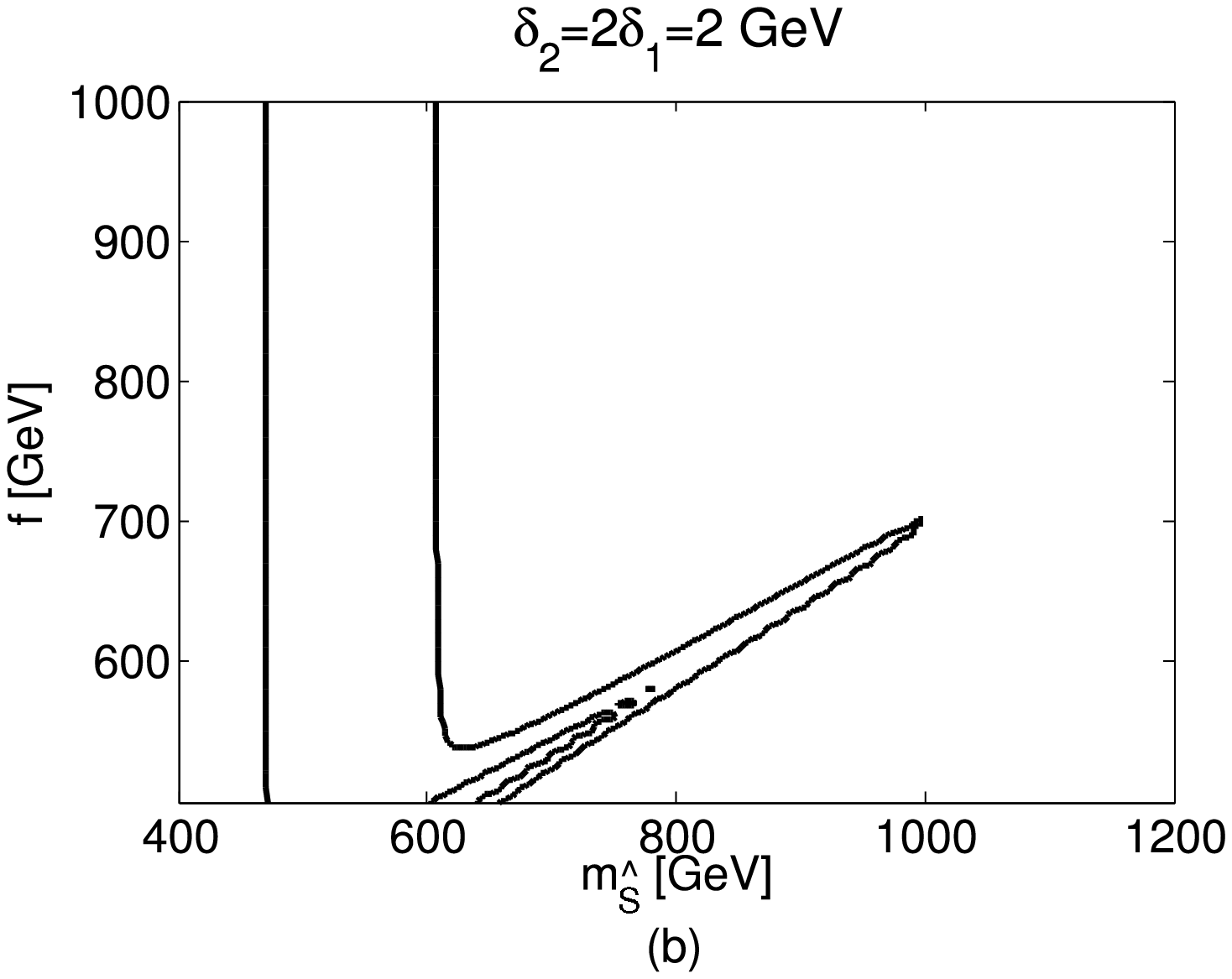}}
\caption{ The left plot (a)  shows the relic density $\Omega h^2$ vs. $m_{\hat{S}}$ for   $f=$ 500 GeV (solid curve), 600 GeV (dashed curve),  and 700 GeV (dash-dotted curve).  The right plot (b) shows the contour plot of the relic density in the $m_{\hat{S}}$ vs. $f$ plane.  
The band in plot (a) and the region enclosed by two contour lines in plot (b) are  the WMAP 3 $\sigma$ regions. The mass splittings $\delta_2=2\delta_1=2$ GeV  is chosen following the LRTH relation.}
\label{fig:relichighf}
\end{center}
\end{figure}

The pole region, which is caused by $\hat{S}\hat{A}\rightarrow{q\bar{q}/l\bar{l}}$ via a heavy gauge boson $Z_H$ exchange,  always appears near $m_{\hat{S}}\sim m_{Z_H}/2$.   The mass of $Z_H$ is determined by the parameter $f$ in the LRTH model.  Changing the value of $f$ would lead to the shift of the pole region, as evident in Fig.~\ref{fig:relichighf}~(a). For larger values of $f$, the pole region shifts to larger values of $m_{\hat{S}}$.   Fig.~\ref{fig:relichighf}~(b) shows the WMAP 3 $\sigma$ region in the $m_{\hat{S}}$ vs. $f$ plane.  The region on the left which is independent of $f$ corresponds to the high mass bulk region I, while the region on the right where $m_{\hat{S}}$ grows with $f$ corresponds to the high mass pole region.

\begin{figure}[bht]
\begin{center}
\resizebox{3.in}{!}{\includegraphics*[83,230][510,562]{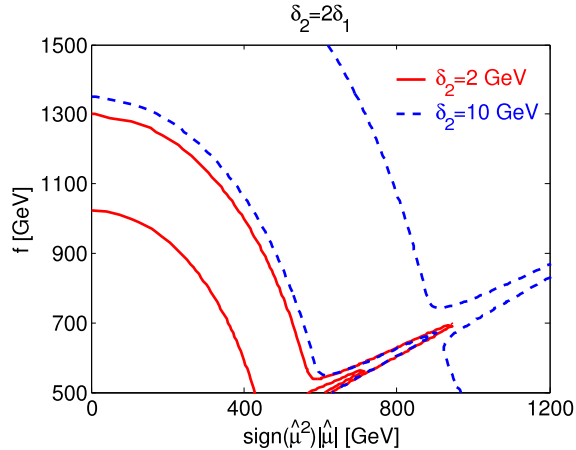}}
\caption{Contour plot for the dark matter relic density  in the $\hat{\mu}$ vs. $f$ plane for the LRTH model.  The mass splitting $\delta_2 = 2 \delta_1=2$ GeV (solid curves) and 10 GeV (dashed curves) are chosen following the LRTH relation. The region enclosed by two contour lines corresponds to the WMAP  3 $\sigma$  region.}
\label{fig:relichighfmuhat}
\end{center}
\end{figure} 

Fig.~\ref{fig:relichighfmuhat} shows the contour plot of the dark matter relic density in the LRTH model parameter space $f$ vs. $\hat\mu$.   For $\hat\mu$=0 GeV, where the mass of the dark matter is mainly provided by the CW potential in the LRTH model, we obtain a region in $f$, $f \gtrsim$ 1000 GeV,  that falls into the WMAP 3 $\sigma$ window.   This is encouraging that with the minimal setup of the LRTH model without introducing an extra mass parameter $\hat\mu^2$ for the dark matter,  it could accommodate the amount of cold dark matter in the Universe.

\section{Conclusion}
\label{sec:conclusion}

The twin Higgs mechanism provides an alternative solution to the little hierarchy problem.  When implemented in the left-right models, there is a natural candidate for WIMP dark matter.  One ${\rm SU}(2)_L$ Higgs doublet couples only to the gauge sector while it does not couple to the SM fermion sector.  The stability of the lightest component in this ${\rm SU}(2)_L$ Higgs doublet is protected by a matter parity.  The dark matter in the LRTH model is similar to that in the IHDM.  However, it has its unique features due to the extra particles in the model.

In this paper, we analyzed the dark matter relic density in the LRTH model when the mass splittings between $\hat{S}$, $\hat{A}$ and $\hat{h}_1^\pm$ follow the approximate relation $\delta_2=2 \delta_1$.  We also generalize our results to the case of the IHDM when such a mass splitting relation is not imposed.  We found that there are two distinctive mass regions  for $m_{\hat{S}}$ where the relic density falls into the WMAP window: (A) low mass region,  and (B) high mass region.  For the low mass region, there are coannihilation pole regions when at least one of $\delta_1$ or $\delta_2$ is small.  There is also a bulk ${WW}$ region around $m_{\hat{S}}\sim$ 70 GeV for small values of $\delta_2$ and a bulk $bb$ region for larger values of $\delta_2$.  Since the relevant particle masses and interactions that entered the relic density calculation also appear in the IHDM, our results also apply to the dark matter relic density in the IHDM.

In the high mass region, there are high mass bulk regions (I) and (II)  due to $\hat{S}\hat{S}\rightarrow {WW/ZZ}$ for $\delta_{1,2}\lesssim$ 15 GeV.  There is also a high mass pole region due to the coannihilation of $\hat{S}$ and $\hat{A}$ via $Z_H$ exchange.  While the pole region only appears in LRTH, the bulk region also exists in the general IHDM. 

In our analysis, we have assumed that the scalar component $\hat{S}$ of $\hat{h}_2^0$ is lighter than the pseudoscalar component $\hat{A}$.  Similar numerical results can be obtained for a pseudoscalar dark matter candidate. 

In summary, there are relatively large regions of parameter space in the LRTH model (as well as in the IHDM) that could accommodate  the amount of cold dark matter in the Universe.  
Even in the minimal setup of the LRTH model with no additional mass parameter introduced for the dark matter, the relic density is still sufficient to close the Universe.
The direct and indirect detection for such a dark matter candidate at current and future dark matter detection experiments can be found in Ref.~\cite{LRTHdetect}.

\begin{acknowledgments}
We would like to thank Z. Chacko and Hock-Seng Goh for useful discussion on the left-right twin Higgs
model.   We would also like to thank A. Pukhov for help with MicrOMEGAs.  
This work is supported under U.S. Department of Energy
contract\# DE-FG02-04ER-41298.

\end{acknowledgments}

\end{document}